\documentclass{ifacconf}

\usepackage{graphicx}      
\usepackage{natbib}        

\usepackage{amsfonts} 
\usepackage{mathtools}
\usepackage{bm, bbm}
\usepackage{xcolor}
\usepackage{comment}

\usepackage{graphicx}
\usepackage{caption}
\usepackage{subcaption}

\newtheorem{theorem}{Theorem}
\newtheorem{corollary}{Corollary}[theorem]

\newtheorem{definition}{Definition}[theorem]
\newtheorem{proposition}{Proposition}[theorem]

\definecolor{ao(english)}{rgb}{0.0, 0.5, 0.0}
\definecolor{americanrose}{rgb}{1.0, 0.01, 0.24}
\definecolor{cerisepink}{rgb}{0.93, 0.23, 0.51}
\definecolor{darkorchid}{rgb}{0.6, 0.2, 0.8}
\definecolor{applegreen}{rgb}{0.55, 0.71, 0.0}
\definecolor{brightpink}{rgb}{1.0, 0.0, 0.5}
\definecolor{azure(colorwheel)}{rgb}{0.0, 0.5, 1.0}
\definecolor{blue-violet}{rgb}{0.54, 0.17, 0.89}
\definecolor{deepmagenta}{rgb}{0.8, 0.0, 0.8}
\definecolor{armygreen}{rgb}{0.29, 0.33, 0.13}
\definecolor{bole}{rgb}{0.47, 0.27, 0.23}

\begin{document}
\begin{frontmatter}

\title{Incentive Design for Large Congestion Games: Publicness-Specific Bayes Correlated Wardrop Equilibrium\\
(Extended Abstract)} 

\thanks[footnoteinfo]{This extended abstract is associated with the presentation in SIAM Annual Meeting 2022 at Pittsburgh, Pennsylvania, U.S.}

\author[First]{Tao Zhang} 

\address[First]{Electrical and Computer Engineering, New York University (e-mail: tz636@nyu.edu).}




\end{frontmatter}

\section{Introduction}

The travel costs of the players (travelers) in anonymous congestion games depend on their choices of routes and also on the states of the transportation network such as incidents, weather, and road work.
In this work, we consider an incomplete-information environment in which the realizations of the states are unobserved by the travelers.
We study how a planner can incentivize the travelers to behave in her favor by strategically designing what and how the travelers get informed about the realizations of the states.

In particular, we study the planner commits to a recommendation mechanism which specifies a joint probability distribution of a profile of mixed-strategy routing policies for the travelers conditioning on the realized state.
The generation of recommendation is a class of the \textit{direct information design} models (see, e.g., \citep{bergemann2016bayes}) in which the travelers are informed about the realizations of the states through the direct recommendation of mixed-strategy routing policies.
In addition, we allow the planner to decide who in the population of travelers receive the same recommendations. 
This is captured by the notion of \textit{publicness}.
In particular, the proposed mechanism contains a \textit{publicness model} which \textit{(i)} determines how many groups the population is partitioned into \textit{(ii)} how many travelers are in each groups.
Given the publicness model, the recommendation rule draws a profile of (mixed-strategy) routing policies and privately sends each routing policy to each group.
Hence, travelers in the same group receives the same recommendation but not the recommendations of other groups.
We refer to such recommendation mechanism as the \textit{publicness-specific} (PS) recommendation mechanism.

We introduce the notion of \textit{publicness-specific Bayes correlated equilibrium} (PS-BCE) as the equilibrium solution concepts of the mechanism design, which extends the concept of Bayes correlated equilibria (BCE) \citep{bergemann2016bayes} by considering the publicness of the recommendations.
The PS recommendation mechanism induces a Bayesian congestion games for the travelers.
We study the limit nonatomic version of such games and prove that the PS-BCE weakly converges to a new equilibrium referred to as \textit{publicness-specific mixed-strategy Bayes correlated Wardrop equilibrium} (PS-BCWE).
Our PS-BCWE extends the concept \textit{Bayes correlated Wardrop equilibrium} (BCWE) \citep{infodesignlarge2021} by taking into account the publicness and considering the travelers' adoption of mixed-strategy routing policies.
Then, we characterize the PS-BCWE and provide some results.

This extended abstract is organized as follows.
In Section \ref{sec:basic_setting}, we describe the basic settings of the congestion game.
Section \ref{sec:PS_mechanism} formulates the PS recommendation mechanism and defines the PS-BCE for the atomic congestion game, while Section \ref{sec:nonatomic+mechanism} formulates the mechanism in nonatomic games and defines the PS-BCWE.
Section \ref{sec:selected_results} show the results regarding the convergence from atomic games to nonatomic games and equilibrium characterizations.

\subsection{Basic Setting: Weighted Atomic Routing Games}\label{sec:basic_setting}

\begin{figure}
  \centering
    \includegraphics[width=\linewidth]{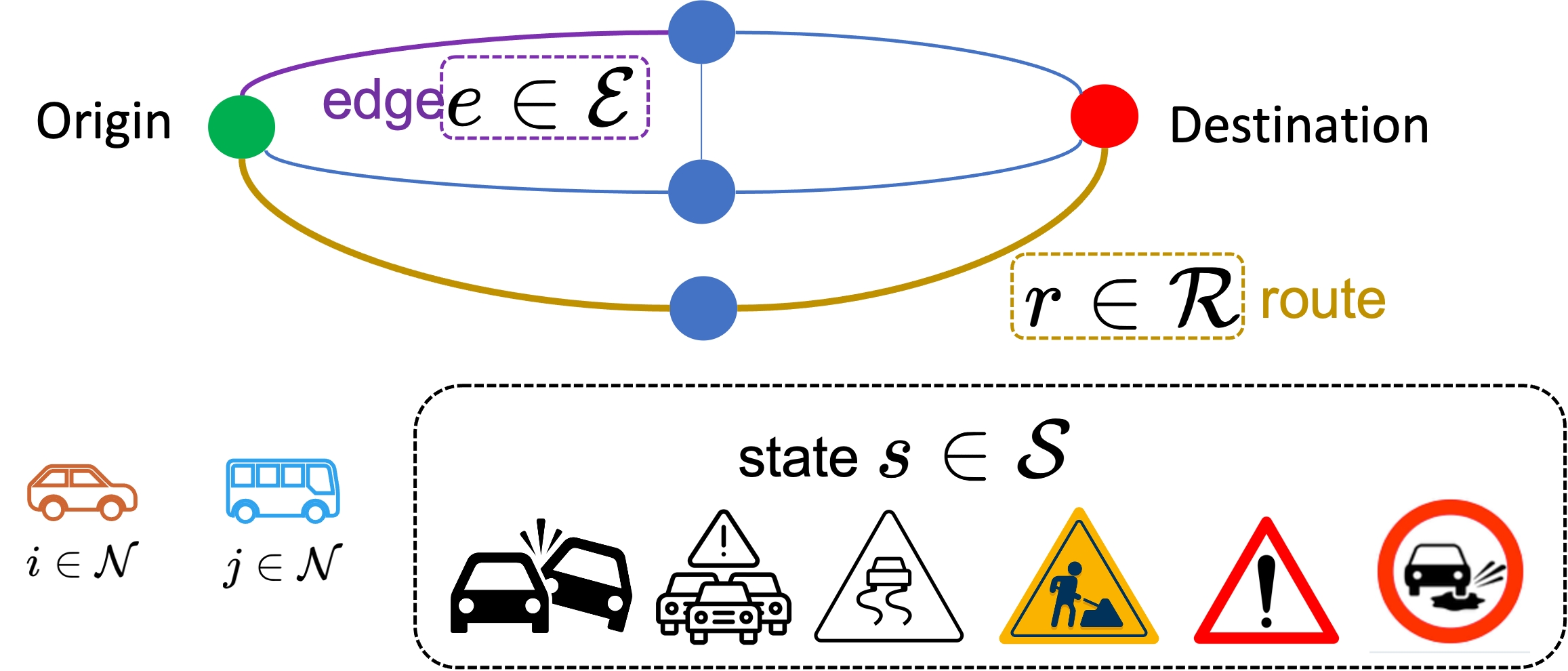}
    \caption{ Direct graph $<\mathcal{R}, \mathcal{E}, \mathcal{S}>$.
    }
    \label{fig:directed_graph}
\end{figure}

With reference to Fig. \ref{fig:directed_graph}, we consider a transportation network modeled as a direct graph, in which there is a finite number of \textit{edges}, denoted as $\mathcal{E}$, a finite number of routes, denoted as $\mathcal{R}$, and a finite number of states, denoted as $\mathcal{S}$.
%
%
Each edge $e\in\mathcal{E}$ is a link that connects a pair of vertices of the graph, and each route $r\in\mathcal{R}$ is a path that connects a pair of origin and destination.
For the ease of exposition, we assume that the network has a single pair of origin and destination.
All our results apply to cases when there are multiple origin-destination pairs.
$\mathcal{S}$ is a finite set of network \textit{states}, which represents a set of possible network conditions, such as incidents, weather, road work, etc.
Each network state $s\in \mathcal{S}$ is randomly drawn by a fictitious player \textit{Nature} from $\mathcal{S}$ according to a probability distribution $p(\cdot)\in\Delta(\mathcal{S})$, which specifies the prior probability of each state.

A \textit{weighted atomic routing game} is defined as a tuple 
\[
\Gamma^{n}_{W}\equiv\big< \mathcal{N}, \mathcal{E}, \mathcal{R}, (\omega_{i})_{i\in \mathcal{N}}, \mathcal{S} \big>,
\]
where $\mathcal{N}\equiv[n]$ is a set of finite travelers with $[n]=(1,2,\dots, n)$.
Each traveler $i\in\mathcal{N}$ has a \textit{weight} $\omega_{i}\in \mathbb{N}_{+}$ and wants to choose a route from the set of routes $\mathcal{R}$.

The \textit{aggregated demand} of the network is given by
\[
D^{n} = \sum\limits_{i\in \mathcal{N}} \omega_{i}.
\]
Given any state $s\in \mathcal{S}$, the decision-makings of travelers induces a \textit{flow} denoted by $f_{r}$ that satisfies the following conditions:
\begin{equation}
    f_{r}\geq 0,
\end{equation}
\begin{equation}
    \sum\limits_{r\in\mathcal{R}} f_{r} = D^{n}.
\end{equation}
A flow profile $f=(f_{r})_{r\in \mathcal{R}}$ uniquely induces an \textit{edge load} for each edge $e\in \mathcal{E}$: 
\begin{equation}
    \ell_{e} = \sum\limits_{r\in\mathcal{R}}f_{r}\mathbf{1}_{\{e\in r\}}.
\end{equation}
When the state is $s$ and the edge load is $\ell_{e}$, the \textit{edge cost} encountered by any traveler traveling on edge $e$ is $C_{e}(\ell_{e}, s)$.
For any edge $e\in \mathcal{E}$ and any state $s\in \mathcal{S}$, the state-dependent \textit{edge cost} function $C_{e}(\cdot, s)$ is a positive, increasing, and differentiable function of edge load.
The state-dependent \textit{route cost} is then given as
\[
C_{r}(f_{r}, f_{-r},s)\equiv \sum\limits_{e\in r} C_{e}(\ell_{e},s).
\]

A \textit{mixed-strategy routing policy profile} is a mapping $\pi:\mathcal{S}\mapsto \prod_{i\in\mathcal{N}}\mathcal{Y}$, where
\[
\mathcal{Y}\equiv\{ y=(y_{r})_{r\in\mathcal{R}}: y_{r}\geq 0, \sum\limits_{r\in\mathcal{R}}y_{r} = 1  \}.
\]
A policy profile $\pi$ can be either \textit{independent} or \textit{correlated} (i.e., joint function).
In the context of routing games, each traveler $i$'s $y_{i}\in \mathcal{Y}$ specifies a probability distribution of traveler $i$'s choices of route or determines the traveler's split of his weight $\omega_{i}$ over $\mathcal{R}$.
Any profile $y=(y_{i}, y_{-i})\in \prod_{i\in\mathcal{N}} \mathcal{Y}$ induces a profile of edge load $\ell(y)=(\ell_{e}(y))_{e\in\mathcal{E}}$.
Suppose that the policy profile $\pi$ draws $y=(y_{i}, y_{-i})$, and each traveler $i$ privately observes $y_{i}=(y_{i,r})_{r\in\mathcal{R}}\in \mathcal{Y}$.
Then, the expected cost when traveler $i$ receives $i$ but chooses route $\hat{y}_{i}\in \mathcal{Y}$ is given as:
\[
\begin{aligned}
&\mathtt{EC}^{n}_{i}(\hat{y}_{i}, y_{i};s)\\
&\equiv \int_{\mathcal{Y}_{-i}} \sum\limits_{r\in\mathcal{R}}\hat{y}_{i,r} \omega^{k}_{i} \sum\limits_{e\in r} C_{e}(\ell_{e}(\hat{y}_{i}, y_{-i}), s) d \pi(y_{i}, y_{-i}|s),
\end{aligned}
\]
where $y_{-i}\equiv \prod_{j\in \mathcal{N}\backslash\{i\}} \mathcal{Y}$.
When $\hat{y}_{i} = y_{i}$, we write $\mathtt{EC}^{n}_{i}( y_{i};s)=\mathtt{EC}^{n}_{i}(y_{i};s)$.

\begin{definition}
A mixed-strategy policy profile $\pi$ is a \textit{correlated equilibrium} if, for all $i\in\mathcal{N}$, $y_{i}, \hat{y}_{i}\in \mathcal{Y}$,
\[
\mathtt{EC}^{n}_{i}(y_{i};s) \leq \mathtt{EC}^{n}_{i}(\hat{y}_{i}, y_{i};s).
\]
\end{definition}

\section{Publicness-Specific Recommendation Mechanism}\label{sec:PS_mechanism}

\begin{figure}
    \centering
    \begin{subfigure}[b]{\linewidth}
    \centering
    \includegraphics[width=\linewidth]{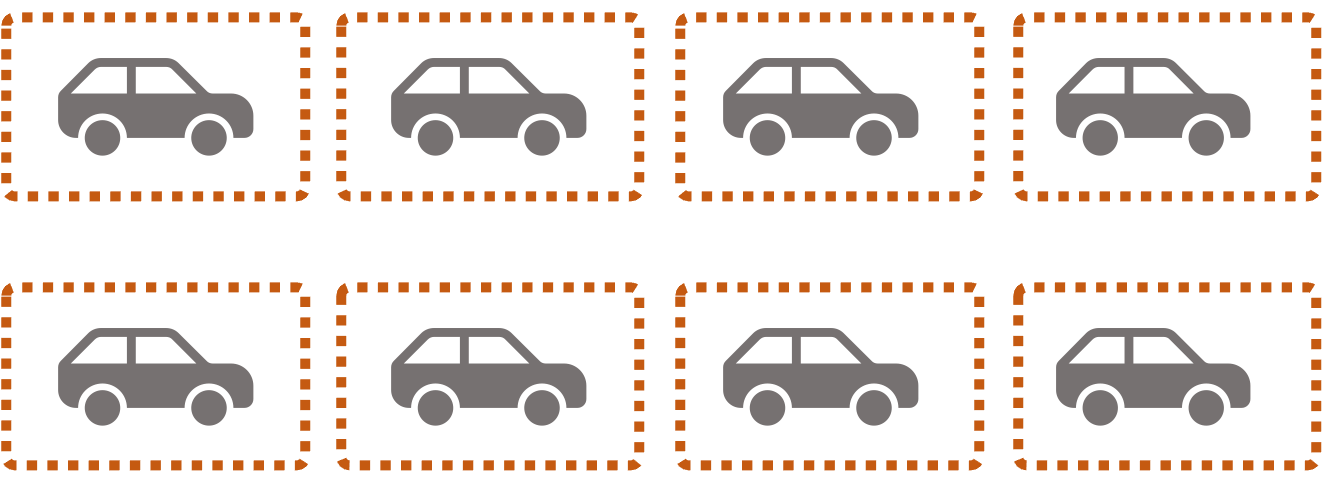}
    \caption{ \label{fig:example_PM_1}}
    \end{subfigure}
    \begin{subfigure}[b]{\linewidth}
    \includegraphics[width=\linewidth]{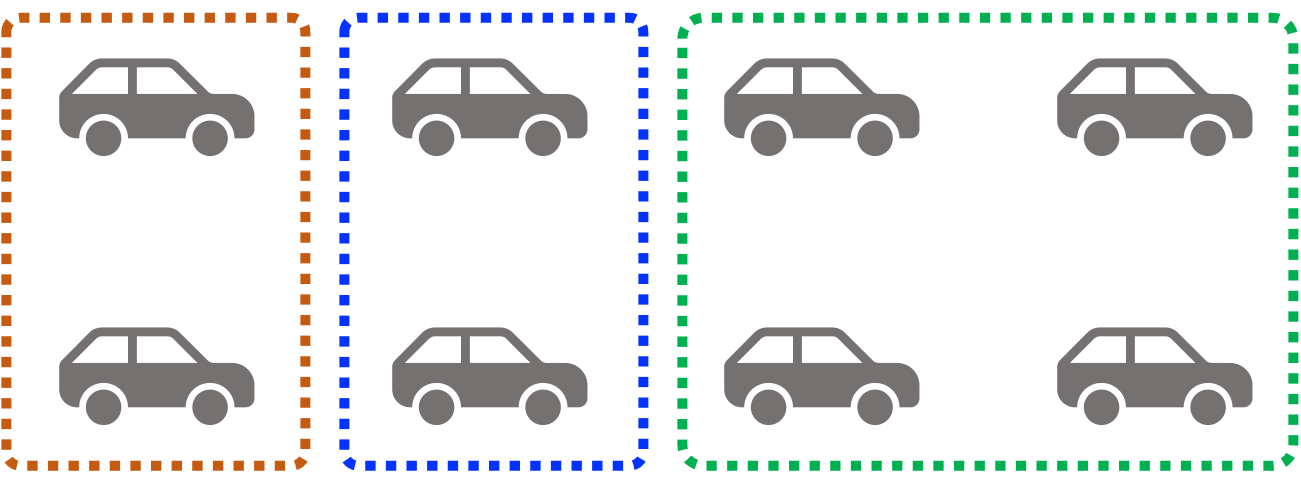}
    \caption{\label{fig:example_PM_3}}
    \end{subfigure}
    \caption{ Two examples of population partitions. In \ref{fig:example_PM_1}, $\mathcal{K}=[8]$, $|\mathcal{N}^{k}|=1$, $x^{k}=\frac{1}{n}$, for all $k\in\mathcal{K}$. In \ref{fig:example_PM_3}, $\mathcal{K}=[3]$, $x^{1}=x^{2}=\frac{1}{4}$, $x^{3}=\frac{1}{2}$.
    }
    \label{fig:PM_example}
\end{figure}

In this section, we consider an incomplete-information environment in which the realizations of the state is unobserved by the travelers.
There is a \textit{central planner} (planner) who privately observes the realizations of the states and provides mixed-strategy routing recommendations (recommendations) for the travelers based on the state.
In particular, the planner offers a \textit{publicness-specific recommendation mechanism} (PS recommendation mechanism), defined by a tuple $<\mathcal{P}^{n,K}, \sigma^{n,K}>$.
Here, $\mathcal{P}^{n,K}\equiv<\mathcal{K}, \{\mathcal{N}^{k}\}_{k\in\mathcal{K}}, \{x^{k}\}_{k\in\mathcal{K}}>$ is the \textit{publicness model}, where
\begin{itemize}
    \item $\mathcal{K}\equiv[K]$ is a set of group indices with $1\leq K\leq n$.
    \item $\mathcal{N}^{k}\subseteq \mathcal{N}$ is the set of travelers in group $k$ with $\prod_{k\in\mathcal{K}}\mathcal{N}^{k} = \mathcal{N}$.
    \item $x^{k}=\frac{|\mathcal{N}^{k}| }{ n }$ is the \textit{partition factor} for group $k$ which represents the portion of travelers of the population $\mathcal{N}$ in group $k$ with $\sum\limits_{k\in \mathcal{K}} x^{k} = 1$. We refer to $\{x^{k}\}_{k\in\mathcal{K}}$ as the \textit{partition profile}. 
\end{itemize}
The publicness model $\mathcal{P}^{n,K}$ determines who receive the same recommendation by partitioning the population of travelers $\mathcal{N}$ into $K$ groups.
Fig. \ref{fig:PM_example} shows two examples of population paritions.
In \ref{fig:example_PM_1}, the population of $8$ travelers is partitioned into $8$ groups in which each group has exactly one traveler. Hence, each group has the same partition factor.
In \ref{fig:example_PM_3}, the population of $8$ travelers is partitioned into $3$ groups and different groups could have different partition factors (i.e., $x^{1}=x^{2}=\frac{1}{4}$ and $x^{3}=\frac{1}{2}$).
Moreover, $\sigma^{n,K}:\mathcal{S}\mapsto \Delta\big( \prod_{k\in\mathcal{K}} \mathcal{Y}  \big)$ is the \textit{recommendation rule}, which specifies the distribution of recommendation profiles of mixed-strategy routing policies.
The timing of the PS recommendation mechanism is as follows:
\begin{itemize}
    \item At the ex-ante stage, the planner commits to a PS recommendation mechanism $<\mathcal{P}^{n,K}, \sigma^{n,K}>$ and publicly discloses it.
    \item \textit{Nature} draws a state $s$ according to $p(\cdot)\in \Delta(\mathcal{S})$ and the planner privately observes $s$.
    \item Based on the state $s$, the planner draws a recommendation profile $y=(y^{k}, y^{-k})\in \prod_{k\in\mathcal{K}} \mathcal{Y}$. Then, the planner privately sends each $y^{k}$ to each group $k$ such that all travelers in group $k$ receives the same recommendation $y^{k}$. Other groups' recommendations $y^{-k}$ remains unobserved by travelers in each group $k$.
    \item Based on the recommendations $y^{k}$, each traveler $i\in\mathcal{N}^{k}$ in each group $k$ chooses a policy $\hat{y}^{k}_{i}\in \mathcal{Y}$.
\end{itemize}

A realized recommendation profile $y=(y^{k}_{r}, y^{-k}_{r})_{r\in\mathcal{R}}$ induces a flow profile $f(y)=(f_{r}(y))_{r\in\mathcal{R}}$, where, for all $r\in\mathcal{R}$,
\[
f_{r}(y) = \sum\limits_{k\in\mathcal{K}} \sum\limits_{i\in \mathcal{N}^{k}} y^{k}_{r}\omega^{k}_{i}.
\]
Given $f(y)$, a profile of edge load $\ell(y)=(\ell_{e}(y))_{e\in\mathcal{E}}$ can be uniquely determined, where, for all $e\in\mathcal{E}$,
\[
\ell_{e}(y) = \sum\limits_{r\ni e} f_{r}(y) = \sum\limits_{r\ni e} \sum\limits_{k\in\mathcal{K}} \sum\limits_{i\in \mathcal{N}^{k}} y^{k}_{r}\omega^{k}_{i}.
\]

Suppose $y=(y^{k}_{r}, y^{-k}_{r})_{r\in\mathcal{R}}$ is recommended but only traveler $i\in\mathcal{N}^{k}$ deviates from the recommendation $y^{k}$ and chooses $\hat{y}^{k}_{i}=(\hat{y}^{k}_{i,r})_{r\in\mathcal{R}}$ instead.
Then, the expected travel cost of this traveler, while all other travelers follow the recommendations, is given as:
\begin{equation}\label{eq:atomic_expected_cost}
    \begin{aligned}
   & \mathtt{EC}^{n,K}_{i}(\hat{y}^{k}_{i}; y^{k}) \equiv \sum\limits_{s\in\mathcal{S}}p(s)\int\limits_{\mathcal{Y}^{-k}} \sum\limits_{r\in\mathcal{R}}\hat{y}^{k}_{i,r}\omega^{k}_{i}\\
   &\times\sum\limits_{e\in r} C_{e}\big( \ell_{e}(\hat{y}^{k}_{i},  y^{k}, y^{-k}),s \big)d \sigma^{n,K}(y^{k}, y^{-k}|s).
    \end{aligned}
\end{equation}
When $\hat{y}^{k}_{i}=y^{k}$, we write $\mathtt{EC}^{n,K}_{i}( y^{k})=\mathtt{EC}^{n,K}_{i}(y^{k}_{i}; y^{k})$, for all $i\in\mathcal{N}^{k}$, $k\in\mathcal{K}$.

For each traveler $i\in\mathcal{N}^{k}$ to have an incentive to follow the planner's recommendation, the planner needs to ensure that the recommendation $y^{k}$ is always preferred to any other mixed-strategy routing policy $\hat{y}^{k}_{i}\in\Delta(\mathcal{Y})$ for every traveler $i\in\mathcal{N}^{k}$ conditional on the traveler's knowledge about the game (i.e., the state and other group's recommendations) through the recommendation $y^{k}$.
Such incentive compatibility is captured by the following notion of obedience.

\begin{definition}\label{def:obedient_atomic}
The PS recommendation mechanism $<\mathcal{P}^{n,K}, \sigma^{n,K}>$ is \textit{obedient} if, for all $i\in\mathcal{N}^{k}$, $k\in\mathcal{K}$, $\hat{y}^{k}_{i}, y^{k}\in \mathcal{Y}$,
\[
\begin{aligned}
\mathtt{EC}^{n,K}_{i}(y^{k}) \leq \mathtt{EC}^{n,K}_{i}(\hat{y}^{k}_{i}; y^{k}).
\end{aligned}
\]
\end{definition}

\begin{definition}\label{def:PS_BCE}
Given a publicness model $\mathcal{P}^{n,K}$, the recommendation rule $\sigma^{n,K}$ constitutes a \textit{publicness-specific Bayes correlated equilibrium} (PS-BCE) if $<\mathcal{P}^{n,K}, \sigma^{n,K}>$ is obedient.
\end{definition}

The planner has an objective function $J(\mathcal{P}^{n,K}, \sigma^{n,K})$ that depends on the probability distribution of the route flows (and edge loads). That is,
\[
J(\mathcal{P}^{n,K}, \sigma^{n,K})\equiv \mathbb{E}^{\sigma^{n,K}}\Big[\sum\limits_{r\in\mathcal{R}} \bar{C}_{r}( \tilde{f}_{r}(\tilde{y} ), \tilde{s})  \Big],
\]
where $\mathcal{E}^{\sigma^{n,K}}[\cdot]$ is the expectation operator under $<\mathcal{P}^{n,K}, \sigma^{n,K}>$ that takes expectation over the flows and the states.
The planner's mechanism design problem is
\begin{equation}
    \begin{aligned}
    \min\limits_{ \mathcal{P}^{n,K}, \sigma^{n,K}  }J(\mathcal{P}^{n,K}, \sigma^{n,K})
    \textup{ s.t., }&  \mathtt{EC}^{n,K}_{i}(y^{k}) \leq \mathtt{EC}^{n,K}_{i}(\hat{y}^{k}_{i}; y^{k}), \\
    &\forall i\in\mathcal{N}^{k}, k\in\mathcal{K}, \hat{y}^{k}_{i}, y^{k}\in \mathcal{Y}.
    \end{aligned}
\end{equation}
If $\bar{C}_{r}(\cdot) = C_{r}(\cdot)$, then the planner aims to minimize the traveling costs for the entire transportation network.

\subsection{Nonatomic PS-recommendation mechanism}\label{sec:nonatomic+mechanism}

In this section, we study PS recommendation mechanism design when the number of travelers goes to infinite; i.e., $n\rightarrow \infty$.
Suppose that the population has a finite demand $D$.
We define a new mechanism $<\mathcal{P}^{K} ,\sigma^{K}>$, where the recommendation rule $\sigma^{K}$ is the same as $\sigma^{n,K}$.
The publicness model becomes $\mathcal{P}^{K}\equiv<\mathcal{K}, \{x^{k}\}_{k\in\mathcal{K}}>$, where $\mathcal{K}=[K]$ is the set of group indices with $1\leq K \leq n$ and the partition profile $\{x^{k}\}_{k\in\mathcal{K}}$ satisfies
\[
x^{k}>0, \sum\limits_{k\in\mathcal{K}} x^{k} = 1.
\]
A partition profile $x=\{x^{k}\}_{k\in\mathcal{K}}$ and a recommendation profile $y=(y^{k}, y^{-k})$ induce profiles of route flows $f(y,x) = (f_{r}(y,x))_{r\in\mathcal{R}}$ and edge loads $\ell(y,x)=(\ell_{e}(y,x))_{e\in\mathcal{E}}$, respectively, where, for all $r\in\mathcal{R}$,
\[
f_{r}(y,x) = \sum\limits_{k\in\mathcal{K}} y^{k}_{r} x^{k} D,
\]
and for all $e\in\mathcal{E}$,
\[
\ell_{e}(y,x) = \sum\limits_{k\in\mathcal{K}} \sum\limits_{r\ni e} f_{r}(y,x) = \sum\limits_{k\in\mathcal{K}} \sum\limits_{r\ni e}y^{k}_{r} x^{k} D.
\]
Suppose that the recommendation rule $\sigma^{K}$ realizes a profile $y=(y^{k}, y^{-k})$, but traveler $i$ in group $K$ (i.e., receiving $y^{k}$) chooses $\hat{y}^{k}_{i}$, while others follow the recommendations.
Then, the traveler's expected travel cost is given as
\begin{equation}\label{eq:cost_nonatomic_psbcwe}
    \begin{aligned}
    &\mathtt{EC}^{K}(\hat{y}^{k}_{i}; y^{k}, x) \equiv \sum\limits_{ s\in\mathcal{S} } p(s) \int\limits_{\mathcal{Y}^{-k}}\\
    &\times\sum\limits_{r\in\mathcal{R}} \hat{y}^{k}_{i,r}\sum\limits_{e\in r} C_{e}(\ell_{e}(y^{k}, y^{-k}, x), s) d \sigma^{K}(y^{k}, y^{-k}|s).
    \end{aligned}
\end{equation}

We extend the notion of obedience in Definition \ref{def:obedient_atomic} to the nonatomic case as follows.
\begin{definition}
The mechanism $<\mathcal{P}^{K}, \sigma^{K}>$ is \textit{obedient} if, for all $k\in\mathcal{K}$, $\hat{y}^{k}_{i}, y^{k}\in\mathcal{Y}$,
\[
\mathtt{EC}^{K}( y^{k}, x) \leq \mathtt{EC}^{K}(\hat{y}^{k}_{i}; y^{k}, x).
\]
\end{definition}

\begin{definition}
Given $\mathcal{P}^{K}$, the recommendation rule $\sigma^{K}$ constitutes a \textit{publicness-specific} \textit{mixed-strategy} \textit{Bayes correlated Wardrop equilibrium} (PS-BCWE) if $<\mathcal{P}^{K}, \sigma^{K}>$ is obedient.
\end{definition}

\subsection{Related Work}

Our work is related to a rapidly growing body of literature in Bayesian persuasion and information design (e.g., \cite{kamenica2011bayesian,rayo2010optimal,ely2017beeps,bergemann2019information,hahn2020prophet,babichenko2021bayesian,celli2020private,mathevet2020information,zhang2021informational,zhang2022bayesian}). Detailed literature review are in the complete version of this work.

In this section, we highlight some existing works that are closely related to our equilibrium concepts.
\cite{bergemann2016bayes} defines the concept of \textit{Bayes correlated equilibria} (BCE) 
The BCE can be used as an equilibrium solution concept for direct information design (i.e., action recommendation) for atomic incomplete-information games.
Our PS-BCE becomes a BCE if $K=n$ and $\mathcal{Y}=\mathcal{R}$ is a set of all pure-strategy policies. 
Then, the expected cost function in (\ref{eq:atomic_expected_cost}) can be rewritten as
\[
\begin{aligned}
   & \mathtt{EC}^{n,n}_{i}(\hat{r}_{i}; r_{i}) \equiv \sum\limits_{s\in\mathcal{S}}p(s)\sum\limits_{r_{-i}} \omega^{k}_{i}\\
   &\times\sum\limits_{e\in r} C_{e}\big( \ell_{e}(\hat{r}_{i},  r_{-i}),s \big) \sigma^{n,K}(r_{i}, r_{-i}|s),
\end{aligned}
\]
where 
\[
\ell_{e}(\hat{r}_{i},  r_{-i}) = \sum\limits_{j\in \mathcal{N}} \omega_{j}\mathbf{1}_{\{e\in r_{j}\}}.
\]
The recommendation rule $\sigma^{n,n}$ is a BCE if for all $i\in\mathcal{N}$, $\hat{r}_{i}, r_{i}\in \mathcal{R}$, 
\[
\mathtt{EC}^{n,n}_{i}(r_{i}; r_{i})\leq \mathtt{EC}^{n,n}_{i}(\hat{r}_{i}; r_{i}).
\]

\cite{infodesignlarge2021} extends the BCE to nonatomic games and defines the concept of \textit{Bayes correlated Wardrop equilibria} (BCWE). 
Our PS-BCWE becomes a mixed-strategy BCWE if $K=1$.
Hence, the expected cost function in (\ref{eq:cost_nonatomic_psbcwe}) becomes
\[
\begin{aligned}
    &\mathtt{EC}^{1}(\hat{y}_{i}; y, x)\\
    & \equiv \sum\limits_{ s\in\mathcal{S} } p(s) \int\limits_{\mathcal{Y}}\sum\limits_{r\in\mathcal{R}} \hat{y}^{k}_{i,r}\sum\limits_{e\in r} C_{e}(\ell_{e}(y, x), s) d \sigma^{K}(y|s).
    \end{aligned}
\]
Then, the recommendation rule $\sigma^{n,1}$ constitutes a mixed-strategy BCWE if for all $i\in\mathcal{N}$, $\hat{y}_{i}, y\in\mathcal{Y}$,
\[
\mathtt{EC}^{1}(y_{i}; y, x)\leq \mathtt{EC}^{1}(\hat{y}_{i}; y, x).
\]


\cite{infodesignlarge2021} also define a multi-population BCWE (MP-BCWE) with fixed population sizes (i.e., demand).
Let $\mathcal{K}=[K]$ be the set of multiple population indexes. Each population $k$ has a demand $\gamma^{k}>0$.
Hence, the total demand of the network is $D=\sum\limits_{k\in\mathcal{K}} \gamma^{k}$.
Equivalently, we can represent $(\gamma^{k})_{k\in\mathcal{K}}$ in terms of partition profile and $D$. That is, each $\gamma^{k} = x^{k}D$, where $x^{k}  = \frac{\gamma^{k}}{D }$.
Since $(\gamma^{k})_{k\in\mathcal{K}}$ is fixed, $x=(x^{k})_{k\in\mathcal{K}}$ is also fixed.
Define, for all $k\in\mathcal{K}$,
\begin{equation}\label{eq:mixed_MP_BCWE}
    \begin{aligned}
\mathcal{Y}^{k}\equiv \big\{y^{k}=(y^{k}_{r})_{r\in\mathcal{R}}: \forall r\in\mathcal{R}, y^{k}_{r}\geq 0, \sum\limits_{r\in\mathcal{R}} y^{k}_{r} = x^{k}D \big\}.
\end{aligned}
\end{equation}
Let $\phi^{K}:\mathcal{S}\mapsto \Delta(\prod_{k\in\mathcal{K}}\mathcal{Y}^{k} )$ denote the recommendation rule that specifies the joint probability distribution, conditioning on the state, of the distribution of route choices (i.e., $y=(y^{k})\in \prod_{k\in\mathcal{K}} \mathcal{Y}^{k}$).
Our PS recommendation mechanism can also use MP-BCWE as the equilibrium solution concept by considering the partition profile $x$ as a decision variable in addition to the recommendation rule $\phi^{K}$.
However, the structure of $\phi^{K}$ directly depends on the choice of he partition profile $x$ due to (\ref{eq:mixed_MP_BCWE}).
In our PS-BCWE, on the other hand, the recommendation rule $\sigma^{K}$ and the partition profile $x$ are independent from each other.
Moreover, every obedience mechanism that is implemented in MP-BCWE can also implemented by our PS-BCWE, but not vice versa.

\section{Selected Results}\label{sec:selected_results}

In this section, we summarize some results of this work.

\subsection{Convergence of $<\mathcal{P}^{n,K}, \sigma^{n,K}>$ to $<\mathcal{P}^{K}, \sigma^{K}>$}

We introduce the following condition.
\begin{equation}\tag{\texttt{CV}}\label{eq:condition_convergence}
    \begin{aligned}
    n\rightarrow \infty,&\\
    \max\limits_{i\in\mathcal{N}} \omega_{i}\stackrel{ n\rightarrow \infty }{\longrightarrow} 0,&\\
    D^{n}\stackrel{ n\rightarrow \infty }{\longrightarrow} D .&
    \end{aligned}
\end{equation}

\begin{theorem}
Under the condition \ref{eq:condition_convergence}, the PS recommendation mechanism $<\mathcal{P}^{n,K}, \sigma^{n,K}>$ weakly converges to $<\mathcal{P}^{K}, \sigma^{K}>$.
\end{theorem}

\subsection{ Bounded Partition Disparity}

In this section, we provide a necessary and sufficient condition for the partition profiles such that the corresponding PS recommendation mechanism is obedient.

By fixing $0<K<\infty$, define the set of \textit{achievable (expected) edge loads:} 
\[
\begin{aligned}
\mathtt{EL}[K]\equiv \big\{ \ell = (\ell_{e})_{e\in\mathcal{E}}: \exists \textup{ obedient } <\mathcal{P}^{K}, \sigma^{K}> \textit{ induces } \ell .\big\}
\end{aligned}
\]
Hence, for every element $\ell^{\dagger}\in \mathtt{EL}[K]$, there is an obedient PS recommendation mechanism $<\mathcal{P}^{K}, \sigma^{K}>$ that induces $\ell^{\dagger}$.
Given any $\ell^{\dagger} \in \mathtt{EL}[K]$, we define the set of \textit{implementing partition profiles:}
\[
\begin{aligned}
\mathtt{IP}[\ell^{\dagger}] \equiv \big\{& x = (x^{k})_{k\in\mathcal{K}}: x^{k}>0, \sum\limits_{k\in\mathcal{K}} x^{k} = 1, \\
&\exists \sigma^{K}, \textup{ s.t., } <\mathcal{P}^{K}, \sigma^{K}> \textit{ achieves } \ell^{\dagger} \big\}.
\end{aligned}
\]
Hence, $\mathtt{IP}[\ell^{\dagger}]$ collects all the partition profiles each of which is a part of an obedient mechanism $<\mathcal{P}^{K}, \sigma^{K}>$ that induces (or achieves) $\ell^{\dagger}$; or, equivalently, every partition profile in $\mathtt{IP}[\ell^{\dagger}]$ implements the edge load $\ell^{\dagger}$.

Any flow profile uniquely induces a profile of edge load, but a profile of edge load may have multiple flow profiles that induce it.
Then, the set of \textit{attainable flow profiles} that induces any $\ell^{\dagger}\in \mathtt{EL}[K]$ is defined as follows:
\[
\begin{aligned}
\mathtt{AF}[\ell^{\dagger}] \equiv \big\{& f = (f_{r})_{r\in\mathcal{R}}: f_{r}\geq 0, \sum\limits_{r\in\mathcal{R}} f_{r} = D, \\
&\sum\limits_{r\ni e} f_{r} = \ell^{\dagger}_{e}, \forall e\in\mathcal{E} \big\}.
\end{aligned}
\]
Therefore, every flow profile in $\mathtt{AF}[\ell^{\dagger}]$ uniquely induces $\ell^{\dagger}$.

Let $x^{\dagger}=(x^{\dagger,k}, x^{\dagger, j}, x^{\dagger-kj})\in \mathtt{IP}[\ell^{\dagger}]$ for any $\ell^{\dagger}\in \mathtt{EL}[K]$, where $x^{\dagger-kj}=(x^{\dagger}_{m})_{m\in \mathcal{K}\backslash\{k,j\}}$.
Given $\ell^{\dagger}\in\mathtt{EL}[K]$ and $x^{\dagger}\in \mathtt{IP}[\ell^{\dagger}]$, a \textit{$kj$-pair-residue attainable flow profile} ($kj$-pra flow profile) is any flow profile $f=(f^{k}_{r}, f^{j}_{r}, f^{-kj}_{r})_{r\in\mathcal{R}}\in \mathtt{AF}[\ell^{\dagger}]$ in which for each $f^{m}=(f^{m}_{r})_{r\in\mathcal{R}}$, $m\in\mathcal{K}\backslash\{k,j\}$, there exists $y^{m}=(y^{m}_{r})_{r\in\mathcal{R}}$ such that $f^{m}_{r} = x^{\dagger,m}y^{m}_{r}D$.
Here, a $kj$-pra flow profile depends on $x^{\dagger}$ only through $x^{\dagger-kj}$ and is independent of $x^{\dagger,k}, x^{\dagger, j}$.
Let the set of $kj$-pra attainable flow as follows: for any $k,j\in\mathcal{K}$,
\[
\begin{aligned}
\mathtt{PR}[\ell^{\dagger}, x^{\dagger-kj}]\equiv \big\{&f = (f^{g}_{r})_{r\in\mathcal{R}}^{g\in\mathcal{K}}\in \mathtt{AF}[\ell^{\dagger}]: \exists y^{m}\in \mathcal{Y}, \\
&\forall m\in\mathcal{K}\backslash\{k,j\}, f^{m}_{r} = x^{\dagger,m}y^{m}_{r}D \big\}.
\end{aligned}
\]
For any $f=(f^{k}_{r}, f^{j}_{r}, f^{-kj}_{r})_{r\in\mathcal{R}}\in \mathtt{PR}[\ell^{\dagger}, x^{\dagger-kj}]$, we refer to $f^{kj}=(f^{k}_{r}, f^{j}_{r})_{r\in\mathcal{R}}$ as the \textit{$kj$-residues} of $f$.
Finally, we define the set of residues for any $\ell^{\dagger}$, $x^{\dagger}=(x^{\dagger,k},x^{\dagger,j},x^{\dagger-kj})\in\mathtt{IP}[\ell^{\dagger}]$, $f^{-kj}=(f^{m}_{r})^{m\neq k,j}_{r\in\mathcal{R}}$:
\[
\begin{aligned}
&\mathtt{F}[f^{-kj};x^{\dagger-kj},\ell^{\dagger}]\equiv\big\{f^{kj}: (f^{kj},f^{-kj})\in \mathtt{PR}[\ell^{\dagger}, x^{\dagger-kj}] \big\}.
\end{aligned}
\]

For any two pairs $f^{kj}=(f^{k},f^{j}),\hat{f}^{kj}=(\hat{f}^{k},\hat{f}^{j})\in \mathtt{F}[f^{-kj}; x^{\dagger-kj}, \ell^{\dagger}]$, we define the $r$-\textit{distance} of $f^{kj}$ and $\hat{f}^{kj}$, for any $r\in\mathcal{R}$,
\begin{equation}
    \begin{aligned}
    \lambda_{r}(f^{kj},\hat{f}^{kj} )\equiv \frac{1}{D}\sum\limits_{g=k,j}(f^{g}_{r} - \hat{f}^{g}_{r} ),
    \end{aligned}
\end{equation}
which is independent of $x^{\dagger,k}$ and $x^{\dagger,j}$.
Let 
\begin{equation}
    \begin{aligned}
    &\Gamma\big(\ell^{\dagger}, x^{\dagger\mid kj}  \big) \equiv (1-\sum\limits_{m\neq k, j}x^{\dagger,m})\\
    &- \min \limits_{f\in \mathtt{PR}[\ell^{\dagger},x^{\dagger-kj}]}\sum\limits_{r\in\mathcal{R}} \max\limits_{\bar{f}^{kj}, \hat{f}^{kj}\in \mathtt{F}[f^{-kj};x^{\dagger-kj}],\ell^{\dagger}]} \lambda_{r}(\bar{f}^{kj}, \hat{f}^{kj}).
    \end{aligned}
\end{equation}
It is clear that $\Gamma\big(\ell^{\dagger}, x^{\dagger\mid kj}  \big)$ is independent of $x^{\dagger,k}$ and $x^{\dagger,j}$.

\begin{definition}
We say that a partition profile $x^{\dagger}\in\mathtt{IP}[\ell^{\dagger}]$ exhibits \textit{bounded partition disparity} (BPD) if, for any pair $k,j\in\mathcal{K}$, 
\begin{equation}
    \begin{aligned}
    0\leq |x^{\dagger,k}-x^{\dagger,j}|\leq \Gamma(\ell^{\dagger}, x^{\dagger-kj}).
    \end{aligned}
\end{equation}
Let $\mathtt{BPD}[\ell^{\dagger}]$ denote a set of partition profiles that exhibits BPD.
\end{definition}

Then, we have the following result.
\begin{theorem}\label{thm:bounded_partition_disparity}
For any achieveable edge load $\ell^{\dagger}\in\mathtt{EL}[K]$, $\mathtt{IP}[\ell^{\dagger}] = \mathtt{BPD}[\ell^{\dagger}]$.
\end{theorem}

Theorem \ref{thm:bounded_partition_disparity} shows that every partition profile that implements the edge load $\ell^{\dagger}$ satisfies the BPD.
That is, the difference between each pair of partition factors, $x^{\dagger,k}$ and $x^{\dagger,j}$, of each $x^{\dagger}\in \mathtt{IP}[\ell^{\dagger}]$ is upper bounded by a term that depends on $x^{\dagger-kj}$ and $\ell^{\dagger}$.
The following corollary directly follows Theorem \ref{thm:bounded_partition_disparity}.

\begin{corollary}
Let $\ell^{\dagger}\in \mathtt{EL}[K]$. Then, the following two hold.
\begin{itemize}
    \item[(i)] If $\Gamma(\ell^{\dagger}, x^{\dagger-kj})=0$ for all $k,j\in\mathcal{K}$, all $x^{\dagger}\in \mathtt{IP}[\ell^{\dagger}]$, then $|\mathtt{IP}[\ell^{\dagger}]|=1$ and each $x^{\dagger,m} = x^{\dagger,m'}$, for all $m,m'\in\mathcal{K}$.
    \item[(ii)] If $\Gamma(\ell^{\dagger}, x^{\dagger-kj})=1$ for all $k,j\in\mathcal{K}$, all $x^{\dagger}\in \mathtt{IP}[\ell^{\dagger}]$, then $\ell^{\dagger}$ is independent of partition profiles.
\end{itemize}
\end{corollary}

\subsection{Characterization of Obedience}

In this section, we show a necessary and sufficient condition for the partition profiles such that the PS recommendation mechanism $<\mathcal{P}^{K}, \sigma^{K}>$ is obedient given \textit{any arbitrary} recommendation rule $\sigma^{K}$.

For any $z^{k}, y^{k}\in \mathcal{Y}$, define a set of \textit{intermediates} connecting $z^{k}$ and $y^{k}$:
\[
\begin{aligned}
\Lambda(z^{k}, y^{k}) \equiv \big\{z^{k} + \delta(y^{k}  - z^{k}): 0\leq \delta \leq 1 \big\}.
\end{aligned}
\]
It is straightforward to verify that $\Lambda(z^{k}, y^{k})\subseteq \mathcal{Y}$ for any $z^{k}, y^{k}\in \mathcal{Y}$.
Let $\alpha:[0,1] \mapsto \mathcal{Y}$ such that $\alpha(0) = z^{k}$, $\alpha(1)=y^{k}$, and $\alpha(\delta)\in \Lambda(z^{k}, y^{k})\backslash\{z^{k}, y^{k}\}$. 
Let $b[T]=(b_{\tau})_{\tau=1}^{T}$ be an ordered sequence (i.e., $0=b_{1}<b_{2}<\cdots<b_{T}=1$) of length $1\leq  T \leq \infty$ in which each $b_{\tau}\in [0,1]$.
Given $b[T]$, define the \textit{path} of length $T$ that connects $z^{k}$ and $y^{k}$ as a vector $P[T]=(P_{\tau})_{\tau=1}^{T}$ where each $P_{\tau} = \alpha(b_{\tau})$, for all $1\leq \tau \leq T$.
Define the set of paths connecting $z^{k}$ and $y^{k}$:
\[
\begin{aligned}
\mathtt{PT}[z^{k}, y^{k}]\equiv \big\{P[T]: b[T]\in \prod_{\tau=1}^{T}[0,1], T\in \mathbb{N}\big\}.
\end{aligned}
\]

Fix any $\sigma^{K}$. Let
\[
\begin{aligned}
L^{\sigma^{K}}(z^{k}, y^{k})\equiv \mathtt{EC}^{K}(z^{k}; z^{k}, x) - \mathtt{EC}^{K}(z^{k}; y^{k}, x).
\end{aligned}
\]
Here, $L^{\sigma^{K}}(z^{k}, y^{k})$ captures the change of the expected travel cost of any individual traveler in group $k$ using policy $z^{k}$ when the planner changes the recommendation from $y^{k}$ to $z^{k}$.
Let
\[
\begin{aligned}
\mathtt{ACL}^{\sigma^{K}}(z^{k}, y^{k})\equiv \inf\limits_{P[T]\in \mathtt{PT}[z^{k}, y^{k}] } \sum\limits_{ \tau=1 }^{T-1} L^{\sigma^{K}}(P_{\tau}, P_{\tau+1}).
\end{aligned}
\]

\begin{proposition}
Fix $\mathcal{K}=[K]$ for any $0<K<\infty$. An arbitrary recommendation rule $\sigma^{K}$ is a PS-BCWE if and only if there exists a partition profile $x=(x^{k})_{k\in\mathcal{K}}$ such that
\begin{itemize}
    \item[(i)] $\mathtt{ACL}^{\sigma^{K}}(z^{k}, y^{k}) + \mathtt{ACL}^{\sigma^{K}}(y^{k}, z^{k})\geq 0$, and
    \item[(ii)] $\mathtt{ACL}^{\sigma^{K}}(z^{k}, y^{k}) = \mathtt{EC}^{K}(y^{k},x) - \mathtt{EC}^{K}(z^{k},x)$.
\end{itemize}
\end{proposition}


\bibliography{ifacconf}             

\begin{thebibliography}{12}
\providecommand{\natexlab}[1]{#1}
\providecommand{\url}[1]{\texttt{#1}}
\providecommand{\urlprefix}{URL }
\expandafter\ifx\csname urlstyle\endcsname\relax
  \providecommand{\doi}[1]{doi:\discretionary{}{}{}#1}\else
  \providecommand{\doi}{doi:\discretionary{}{}{}\begingroup
  \urlstyle{rm}\Url}\fi

\bibitem[{Babichenko et~al.(2021)Babichenko, Talgam-Cohen, and
  Zabarnyi}]{babichenko2021bayesian}
Babichenko, Y., Talgam-Cohen, I., and Zabarnyi, K. (2021).
\newblock Bayesian persuasion under ex ante and ex post constraints.
\newblock In \emph{Proceedings of the AAAI Conference on Artificial
  Intelligence}, volume~35, 5127--5134.

\bibitem[{Bergemann and Morris(2016)}]{bergemann2016bayes}
Bergemann, D. and Morris, S. (2016).
\newblock Bayes correlated equilibrium and the comparison of information
  structures in games.
\newblock \emph{Theoretical Economics}, 11(2), 487--522.

\bibitem[{Bergemann and Morris(2019)}]{bergemann2019information}
Bergemann, D. and Morris, S. (2019).
\newblock Information design: A unified perspective.
\newblock \emph{Journal of Economic Literature}, 57(1), 44--95.

\bibitem[{Celli et~al.(2020)Celli, Coniglio, and Gatti}]{celli2020private}
Celli, A., Coniglio, S., and Gatti, N. (2020).
\newblock Private bayesian persuasion with sequential games.
\newblock In \emph{Proceedings of the AAAI Conference on Artificial
  Intelligence}, volume~34, 1886--1893.

\bibitem[{Ely(2017)}]{ely2017beeps}
Ely, J.C. (2017).
\newblock Beeps.
\newblock \emph{American Economic Review}, 107(1), 31--53.

\bibitem[{Hahn et~al.(2020)Hahn, Hoefer, and Smorodinsky}]{hahn2020prophet}
Hahn, N., Hoefer, M., and Smorodinsky, R. (2020).
\newblock Prophet inequalities for bayesian persuasion.
\newblock In \emph{IJCAI}, 175--181.

\bibitem[{Kamenica and Gentzkow(2011)}]{kamenica2011bayesian}
Kamenica, E. and Gentzkow, M. (2011).
\newblock Bayesian persuasion.
\newblock \emph{American Economic Review}, 101(6), 2590--2615.

\bibitem[{Koessler et~al.(2021)Koessler, Scarsini, and
  Tomala}]{infodesignlarge2021}
Koessler, F., Scarsini, M., and Tomala, T. (2021).
\newblock Information design in large games.
\newblock \emph{CoRR}, abs/2107.06312.
\newblock \urlprefix\url{https://arxiv.org/abs/2107.06312}.

\bibitem[{Mathevet et~al.(2020)Mathevet, Perego, and
  Taneva}]{mathevet2020information}
Mathevet, L., Perego, J., and Taneva, I. (2020).
\newblock On information design in games.
\newblock \emph{Journal of Political Economy}, 128(4), 1370--1404.

\bibitem[{Rayo and Segal(2010)}]{rayo2010optimal}
Rayo, L. and Segal, I. (2010).
\newblock Optimal information disclosure.
\newblock \emph{Journal of political Economy}, 118(5), 949--987.

\bibitem[{Zhang and Zhu(2021)}]{zhang2021informational}
Zhang, T. and Zhu, Q. (2021).
\newblock Informational design of dynamic multi-agent system.
\newblock \emph{arXiv preprint arXiv:2105.03052}.

\bibitem[{Zhang and Zhu(2022)}]{zhang2022bayesian}
Zhang, T. and Zhu, Q. (2022).
\newblock Bayesian promised persuasion: Dynamic forward-looking multiagent
  delegation with informational burning.
\newblock \emph{arXiv preprint arXiv:2201.06081}.

\end{thebibliography}
                                                   







\end{document}